\documentclass[dvipdfmx,11pt,a4]{article}
\usepackage[top=20truemm,bottom=20truemm,left=20truemm,right=20truemm]{geometry}
\usepackage[dvipdfmx]{graphicx,color}
\usepackage{amssymb,amsmath}
\usepackage{hyperref}
\hypersetup{% hyperrefオプションリスト
setpagesize=false,
 bookmarksnumbered=true,%
 bookmarksopen=true,%
 colorlinks=true,%
 linkcolor=blue,
 citecolor=red,
}
\newcommand{\eq}[1]{\mbox{Eq.~(\ref{#1})}}
\newcommand{\fig}[1]{\mbox{Fig.~\ref{#1}}}
\newcommand{\D}{{\rm d}}
\newcommand\ringring[1]{%
  {% make an Ord atom
   \mathop{\kern0pt #1}\limits^{% set a box over the variable
     \vbox to-1.85ex{
       \kern-2ex % lower the ring accents
       \hbox to 0pt{\hss\normalfont\kern.1em \r{}\kern-.45em \r{}\hss}%
       \vss % fill
     }% end of \vbox
   }% end of the superscript
  }% end of \mathop
}

\begin{document}

%\begin{titlepage}
\vfill
\begin{flushright}
KEK-Cosmo-???,
KEK-TH-????,
RUP-18-27
\end{flushright}

\begin{center}
{\Large{\bf Effect of Inhomogeneity on Primordial Black Hole \\ Formation  in the Matter Dominated Era}}
\vskip 1em
{\bf
Takafumi Kokubu$^{1,2, \ast}$, 
Koutarou Kyutoku$^{1, 3, 4, 5, \ddagger}$,
Kazunori Kohri$^{1, 3, 6, \S}$,
Tomohiro Harada$^{2, \dagger}$}
 \\ 
{\it $^1$ Theory Center, Institute of Particle and Nuclear Studies, KEK,
Tsukuba 305-0801, Japan}\\
{\it $^2$Department of Physics, Rikkyo University, Toshima, Tokyo 171-8501, Japan}\\
{\it $^3$Department of Particle and Nuclear Physics, the Graduate University
for Advanced Studies }\\
{\it (Sokendai), Tsukuba 305-0801, Japan}\\
{\it $^4$ Interdisciplinary Theoretical and Mathematical Sciences Program
(iTHEMS), RIKEN,}\\
{\it Wako, Saitama 351-0198, Japan}\\
{\it $^5$ Center for Gravitational Physics, Yukawa Institute for Theoretical 
Physics, }\\
{\it Kyoto University, Kyoto 606-8502, Japan}\\
{\it $^6$ Rudolf Peierls Centre for Theoretical Physics, The University of Oxford, }\\
{\it 1 Keble Road, Oxford, OX1 3NP, UK}\\
\vskip 1em

$^\ast$ kokubu@post.kek.jp, 
$^\ddagger$ kyutoku@post.kek.jp, %\\
$^\S$ kohri@post.kek.jp,
$^\dagger$ harada@rikkyo.ac.jp
\vskip 1em
{\bf Abstract}
\end{center}
\vskip -0.5em
We investigate the effect of inhomogeneity on primordial black hole formation in the matter dominated era.
In the gravitational collapse of an inhomogeneous density distribution, a black hole forms if apparent horizon prevents information of the central region of the configuration from leaking.
Since information cannot propagate faster than the speed of light, we identify the threshold of the black hole formation by considering the finite speed for propagation of information. 
We show that the production probability $\beta_{inhom}(\sigma)$ of primordial black holes, where $\sigma$ is density fluctuation at horizon entry, is significantly enhanced from that derived in previous work in which the speed of propagation was effectively regarded as infinite. 
For $\sigma \ll 1$, we obtain $\beta_{inhom}\simeq 3.70 \sigma^{3/2}$, which is larger by about an order of magnitude than the probability derived in earlier work by assuming instantaneous propagation of information.

\tableofcontents
%%%%%%%%%%%%%%%%%%%%%%%%%%%%%%%%%%%%%%%%%
%%%SECTION%%%SECTION%%%SECTION%%%SECTION%%%SECTION%
%%%%%%%%%%%%%%%%%%%%%%%%%%%%%%%%%%%%%%%%%
\section{Introduction}\label{Introduction-section}
Primordial black holes (PBHs) are black holes which may have formed in the early universe. Masses of such black holes can take quite a broad range, e.g., they can be smaller than the mass of the Sun \cite{ZeldovichNovikov1967, Hawking1971, CarrHawking1974, Carr1975}.
PBHs can be used as probes into gravitational collapse, the early Universe, high energy physics, dark matter, gravitational waves and quantum gravity (see Refs.~\cite{Carr2005review, Khlopov2010review,  CarrKohriSendoudaYokoyama2010, CarrKuhnelSandstad2016} for reviews). 

Most studies on PBHs have focused attention on the formation in the
radiation dominated era, see, e.g., Ref.~\cite{HaradaYooKohri2013} and
references therein. On the other hand however, PBHs can be also formed
in matter dominated eras, which have also been intensively
investigated recently
\cite{HaradaYooKohriNakaoJhingan2016,HaradaYooKohriNakao2017, KhlopovPolnarev1980,PolnarevKhlopov1981}.

In addition to the well-known matter dominated era after the matter-radiation equality in the standard big-bang cosmology, pressure-less matters
could have dominated the density of the Universe in some periods of
the cosmological history earlier than the Big-bang nucleosynthesis
(BBN) epoch, e.g., in scenarios beyond the standard model.  For
example, in modern scenarios of inflationary cosmology \cite{Lyth:2009zz}, we naturally
expect a matter dominated era by the oscillating inflaton field which
had induced inflation. The oscillation era of the inflaton field
follows the end of inflation until the reheating of the Universe. The
Universe was reheated through thermalization processes of daughter
particles produced by decays of inflaton field. The energy density of
such a scalar field oscillating at its vacuum expectation value with
its mass term behaves like pressure-less matter. Then, the early
matter dominated era is realized soon after the inflaton field starts
its oscillation.  In addition to the inflaton field, more concretely
in supergravity or superstring theory which is expected to be a
candidate of unified theories beyond the standard model, there exist a
lot of massive long-lived scalar fields such as moduli or
dilatons \cite{Banks:1993en, deCarlos:1993wie, Banks:1995dt}. Their lifetime tends to be quite long due to their decays only through gravitational interaction. Then, the early matter dominated era by the oscillation energies of these fields can continue until
just before the BBN epoch with its minimally-possible mass of the
order of weak scales. 

Then, it is a crucial subject to investigate a mechanism of PBH formation
in this kind of the early matter dominated era in the scenarios
beyond the standard model. In this paper we recast possible changes of
the constraints on the amount of PBHs, which come from differences of
the formation rates between radiation and matter domination eras. In
addition, the existence of this early matter dominated era changes a
relation between the mass of the PBHs and the wave number
$k$~\cite{AlabidiKohri2009} of the fluctuation at which the horizon
crosses. These effects can be commonly treated by introducing a new
parameter, $T_R$, which is the reheating temperature after the end of
the early matter domination~\cite{Carr:2017edp,Kohri:2018qtx}.

One may speculate that, unless baryonic processes play a significant role at small scales, any density perturbations in the matter-dominated era lead to formation of a black hole because of the pressure-less property of dust. Such a speculation however needs the following revisions:
PBH formation in matter dominated era is governed by the three effects, namely, anisotropy, spin and inhomogeneity of density perturbations.

For the cases that the density fluctuation $\sigma$ at horizon entry is sufficiently small, the effect of anisotropy of density perturbations have been investigated. Ref. \cite{HaradaYooKohriNakaoJhingan2016} reported that nearly spherical density perturbations can form PBHs.
On the other hand,  aspherical collapse of perturbations results in two-dimensional ``pancake" singularity and then if the particle picture is adopted these perturbations are virialised which prevents them from becoming black holes.

The effect of the spin on PBH formation was investigated in Ref.~\cite{HaradaYooKohriNakao2017}. A rotating perturbation results in a rapidly rotating PBH if the period of matter dominanted era is long enough.
The PBH production rate in the matter dominated era is larger than that in the radiation dominated era for $\sigma \lesssim 0.05$, while they are comparable for $0.05 \lesssim \sigma \lesssim1$.

The effect of inhomogeneity was investigated by Khlopov and Polnarev \cite{KhlopovPolnarev1980, PolnarevKhlopov1981}.
As mentioned, only nearly spherical perturbations can become black holes when $\sigma$ is small enough. 
For spherically symmetric density distributions, the effect of inhomogeneity can be investigated by taking an exact solution to the Einstein equation known as Lema\^itre-Tolman-Bondi (LTB) spacetime.
This spacetime is a general solution describing spherically symmetric inhomogeneous dust collapse. Since this solution is exact, we can follow time evolutions of inhomogeneous dust in a non-perturbative manner.
The LTB solution has also been used as a model of PBH in Refs.~\cite{Harada:2001kc,Harada:2015ewt}.

In the LTB spacetime, from physically reasonable initial data, an apparent horizon forms in a collapsing mass. Moreover, falling dust particles lead to forming a curvature singularity at the center in a finite time.
The two features, the horizon formation and the singularity formation, determine the final state of the collapsing mass:
If the horizon is formed ``earlier than'' the singularity, the collapsing mass becomes a black hole. Otherwise, if the classical picture of general relativity is applicable even at the central region, {\it a naked singularity} will form at the central region, which is not covered by a horizon.  
By comparing the time of the horizon formation and the singularity formation, one can identify a threshold for black hole formation. That was the basic idea proposed by Polnarev and Khlopov in Ref. \cite{PolnarevKhlopov1981} to investigate PBH formation in the matter dominated era.

What really occurs in the central region is unclear.
One possibility is the formation of a  singularity and it is inevitable as far as the LTB solution is adopted.
Another possibility is a change of the property of falling dust as suggested by Polnarev and Khlopov. 
In their scenario the nature of the particles changes from dust to radiation because the velocity dispersion of particles become large when particles are condensed at the central region.
They insisted that the gravitational collapse was stopped by the pressure of radiation and the collapsing particles did not become a black hole.
They also mentioned a scenario of a dispersion of dust particles, which prevents a black hole formation if the particle picture is adopted. When the particles fall into the central region, they pass through each other and disperse again. This process also could prevent a density distribution from forming a black hole.
As mentioned above, the result of gravitational collapse would have many possibilities in the vicinity of the center.
But whatever happens at the center, if that information is hidden behind the horizon before information reaches us, the collapsing matter will become a black hole.
In this article, we collectively call these possibilities (naked singularity, virialized configuration, radiation or dispersion of the dust particles, etc.) the naked singularity.
%Whether a configuration becomes a black hole is determined by comparing the time when the apparent horizon is formed and the time when the central region becomes singular. 
%Therefore, Polnarev and Khlopov have adopted a condition of PBH formation such that the configuration would be a black hole if the apparent horizon is formed ``earlier than'' the central region becomes singular.

Although Polnarev and Khlopov have given the condition of black hole formation, it does not take the propagation of information into account. 
We emphasize that we should treat the notion of ``earlier than''  more carefully;
to compare the time of two spatially distant events, the duration of causal propagation between the events must be taken into account.
To treat black hole formation appropriately, we must consider the time of information propagating from the center to a characteristic radius of the configuration. Since any information propagates at most with the speed of light, the lower bound on the probability of PBH formation is obtained by tracking the null geodesic emanating from the central region to the characteristic radius.
In this paper, we derive a new criterion of PBH formation with taking the propagation of information along null geodesics into account.

The paper is organized as follows:
In Sec. \ref{LTB-section}, relativistic treatment of inhomogeneous dust configuration is introduced. 
We also review previous study on PBH formation in the matter dominated era.
In Sec. \ref{PBH-formation-section}, we investigate null geodesic in LTB spacetime to derive an improved criterion for PBH formation.
Sec. \ref{PBH Production Probability} gives PBH production probability.
Sec. \ref{Conclusion-section} is dedicated to summary and discussions.
We adopt $c=G=1$ unit and diag$(-,+,+,+)$ signature of the metric throughout the paper.

%%%%%%%%%%%%%%%%%%%%%%%%%%%%%%%%%%%%%%%%%
%%%SECTION%%%SECTION%%%SECTION%%%SECTION%%%SECTION%
%%%%%%%%%%%%%%%%%%%%%%%%%%%%%%%%%%%%%%%%%
\section{Collapse of inhomogeneous density distribution}\label{LTB-section}
In this section we introduce the LTB spacetime, an exact solution of the Einstein equation, which can describe gravitationally-collapsing inhomogeneous dust sphere. 
By using this model, we reinterpret the criterion of PBH formation originally derived in Ref. \cite{PolnarevKhlopov1981}. We emphasize that our derivation does not depend on the specific form of the initial density profile.

\subsection{Lema\^itre-Tolman-Bondi spacetime}
The collapse of a spherically symmetric dust configuration is described by the LTB spacetime \cite{Lemaitre1933, Tolman1934, Bondi1947},
\begin{align}
\D s^2=-\D t^2+\frac{(R')^2}{1+f(r)}\D r^2+R^2 (\D \theta^2+\sin^2\theta \D \phi^2), \label{LTB-metric}
\end{align}
where $':=\partial/\partial r$, $f(r)$ is a function determined by the initial condition and $R:=R(t,r)$ is the areal radius of a sphere at the time $t$ and the coordinate $r$.
The stress-energy tensor for the dust is given by, $T^{\mu \nu}=\rho(t,r)u^\mu u^\nu$, where $\rho(t,r)$ is the energy density of the dust and $u^\mu$ is its four velocity.
Greek indices run over $t, r, \theta$ and $\phi$.
The metric and stress-energy tensor give us non-trivial components of the Einstein equations as 
\begin{align}
&\left\{R(-f+\dot R^2) \right\}'=8\pi\rho(t,r)R^2R',  \label{eom1-LTB}\\
&(R\dot R^2)\dot{}=f\dot R, \label{eom2-LTB}
\end{align}
where $\dot{} :=\partial/\partial t$.
By integrating \eq{eom2-LTB} with respect to $t$, we obtain 
\begin{align}
\dot R^2=f(r)+\frac{R_g(r)}{R}, \label{integrated-eom2}
\end{align}
where $R_g(r)$ is an arbitrary function of $r$ at this stage.
By substituting \eq{integrated-eom2} into \eq{eom1-LTB}, we obtain 
\begin{align}
R_g(r)'=8\pi \rho(t,r)R^2R'. \label{deriv-mass}
\end{align}
Since the left hand side of \eq{deriv-mass} is independent of $t$, the right hand side is also time independent, i.e., $8\pi \rho(t,r)R^2R'$ is conserved in time. 
By imposing $R_g(0)=0$, integration  of \eq{deriv-mass} yields
\begin{align}
R_g(r)=\int^{R(t_i,r)}_0 8\pi \rho(t_i, r) R^2(t_i, r) \D R(t_i, r) \label{mass}
\end{align}
with $t_i$ being the initial time when the collapse begins. 
Through the definition of Misner-Sharp mass $m$ \cite{MisnerSharp1964}, we obtain
\begin{align}
2m:=R(1-g^{\alpha\beta}\nabla_\alpha R \nabla_\beta R)=R_g(r),
 \label{misner-sharp}
\end{align}
where $\nabla_\alpha$ is the covariant derivative. 
The last equality in \eq{misner-sharp} is obtained through \eq{integrated-eom2}. 
Equation (\ref{misner-sharp}) shows that $R_g(r)$ is equal to the gravitational radius of the dust contained in $r$.

Time evolution of the areal radius $R(t,r)$ of each shell is determined implicitly by integrating \eq{integrated-eom2} as
\begin{align}
t-t_s(r)=-\sqrt{\frac{R(t,r)^3}{R_g(r)}}F\left[-\frac{f(r)R(t,r)}{R_g(r)}\right]. \label{EOM}
\end{align}
Here, $t_s(r)$ is the trajectory of a singularity, at which $R(t_s(r),r)$ becomes zero and Kretschmann's scalar-invariant $R_{\alpha\beta\gamma\sigma}R^{\alpha\beta\gamma\sigma}$ diverges. 
The general form of $F(y)$ is given in Ref. \cite{JoshiDwivedi1993}.
We consider the configuration which begins to collapse at  the initial time, i.e., $\dot R(t=t_i,r)=0$. Then, $F(y)$ is given by
\begin{align}
F(y)=\frac{\sin^{-1}\sqrt{y}}{y^{3/2}}-\frac{\sqrt{1-y}}{y} \qquad (1\geq y>0).
\label{F(y)}
\end{align}

\subsection{Gravitational collapse}
Let us consider the collapse of a density distribution in full general relativity using the LTB solution.
Without loss of generality we can set the initial time $t_i$ to be zero. We also rescale the radial coordinate as $R(0,r)\equiv r$. From these settings and \eq{integrated-eom2}, we identify $f$ as 
\begin{align}
f(r)=-\frac{R_g(r)}{r}. \label{frRg}
\end{align}
Then the equation of motion reduces to
\begin{align}
t-t_s(r)=-\sqrt{\frac{R(t,r)^3}{R_g(r)}}F\left(\frac{R(t,r)}{r}\right). \label{EOM-bound}
\end{align}
By setting $t=0$ in \eq{EOM-bound}, we first derive the trajectory of singularity $t_s(r)$ as
\begin{align}
t_s(r)=\frac{\pi}{2}\sqrt{\frac{r^3}{R_g(r)}}. \label{tsr-momentarily-static}
\end{align}

Next, we investigate behavior of the apparent horizon. Let us denote the time when the shell with $r$ becomes trapped as $t_{ah}(r)$. 
The apparent horizon is identified as the location where $\nabla^\alpha R \nabla_\alpha R=0$, i.e., the place where the constant $R$ surface becomes null \cite{MiyamotoJhinganHarada2013}.
The trajectory of the apparent horizon is obtained by substituting $R_g(r)=R(t_{ah}(r), r)$ obtained by  \eq{misner-sharp} into \eq{EOM-bound} as 
\begin{align}
t_{ah}(r)-t_s(r)=-R_g(r)F\left(\frac{R_g(r)}{r}\right). \label{ap-horizon}
\end{align}
To simplify equations, we introduce a parameter $x$,
\begin{align}
x:=\frac{R_g(r_1)}{r_1} \label{def-x}
\end{align}
where $r_1$ is the radius of the surface of the confiuguration (we will introduce $r_1$ again in section \ref{PBH-formation-section} with slightly modified definition)
and the mean density of the initial matter distribution $\bar \rho(0,r)$ inside $r$,
\begin{align}
\bar \rho(0,r):=\frac{m}{4\pi r^3/3}=\frac{3R_g(r)}{8\pi r^3}. \label{def-rho-bar}
\end{align}
Since $x$ is the ratio of gravitational radius of the configuration to its radius at the initial state, it characterizes the compactness of the configuration. 
\eq{def-x} and \eq{def-rho-bar} are combined to yield
\begin{align}
x=\frac{8\pi \bar \rho(0, r_1)}{3}r_1^2.
\end{align}
By using new parameters, we recast \eq{tsr-momentarily-static} and \eq{ap-horizon} evaluated at $r=r_1$ as
\begin{align}
t_s(r)&=\sqrt{\frac{3}{8\pi \bar \rho(0,r)}}\frac{\pi}{2}, \label{tsr-variant} \\
t_{ah}(r_1)&=\sqrt{\frac{3}{8\pi \bar \rho(0,r_1)}}\left(\frac{\pi}{2}-F(x)x^{3/2}\right). \label{tah-variant}
\end{align}

\subsection{Criterion of PBH formation due to Polnarev-Khlopov}\label{PK-condition-section}
Based on an argument using LTB spacetime, Polnarev and Khlopov derived a criterion of the PBH formation \cite{PolnarevKhlopov1981}.
They assumed that the PBH is formed if 
\begin{align}
t_{ah}(r_1)<t_s(0). \label{PK-condition}
\end{align}
is satisfied.
According to Ref. \cite{PolnarevKhlopov1981}, 
when the radius $r_1$ of a configuration is much larger than its gravitational radius $R_g(r_1)$, i.e., $x \ll 1$,
\eq{PK-condition} reduces to the following inequality: 
\begin{align}
u\lesssim x^{3/2}, \label{PK-condition-ux}
\end{align}
where $u$ is inhomogeneity of a density distribution defined by
\begin{align}
u:=\frac{\rho(0,0)-\bar \rho(0,r_1)}{\bar \rho(0,r_1)}=\frac{\rho(0,0)}{\bar \rho(0,r_1)}-1. \label{inhomogeneity-u}
\end{align}

We show that the proportionality coefficient of $x^{3/2}$ in \eq{PK-condition-ux} can be determined without identifying the form of the density profile.
By substituting \eq{tsr-variant} and \eq{tah-variant} into the inequality (\ref{PK-condition}), we obtain
\begin{align}
u<f_1(x), \label{PK-condition-variant-variant}
\end{align}
where
\begin{align}
f_1(x):=\left(1-\frac{2}{\pi}F(x)x^{3/2}\right)^{-2}-1.
\end{align}
\eq{PK-condition-variant-variant} is the criterion of PBH formation due to the inequality \eq{PK-condition} by Khlopov and Polnarev.
For $x\ll 1$, \eq{PK-condition-variant-variant} reduces to
\begin{align}
u\lesssim \frac{8}{3\pi}x^{3/2}=0.849x^{3/2}. \label{PK-condition-approx}
\end{align}
$u \ll 1$ is required for this inequality to be satisfied, i.e., only nearly homogeneous configurations can collapse to black holes.
If \eq{PK-condition-approx} is not satisfied, the gravitational collapse results in naked singularities.
The power $x^{3/2}$ in \eq{PK-condition-approx} agrees with \eq{PK-condition-ux} derived in Ref. \cite{PolnarevKhlopov1981}, and the proportionality coefficient is derived here for the first time.
 \eq{PK-condition-approx} is the criterion of PBH formation in the case where the finite speed of information is not considered.
We will point out insufficiency of the criterion of \eq{PK-condition-approx} in the next section.

%%%%%%%%%%%%%%%%%%%%%%%%%%%%%%%%%%%%%%%%%
%%%SECTION%%%SECTION%%%SECTION%%%SECTION%%%SECTION%
%%%%%%%%%%%%%%%%%%%%%%%%%%%%%%%%%%%%%%%%%
\section{Improved criterion of PBH formation}\label{PBH-formation-section}
In this section we obtain an outgoing radial null geodesic propagating from the central singularity to the surface of a density distribution.
Polnarev and Khlopov adopted the condition of PBH formation,  \eq{PK-condition}, that compares two characteristic times on a collapsing configuration. However, this condition misses an important case which would contribute to PBH formation:
Since this condition compares two spatially distant points in a specific coordinate time, {\it the case where an apparent horizon is formed before the signal from the central singularity reaches the surface of the configuration is excluded and the rate of PBH formation is underestimated.}
Hence we derive an improved criterion including the effect of the signal speed. 

In this calculation we take the initial density profile to be 
\begin{align}
\rho=\rho_0\left[1-\left(\frac{r}{r_*} \right)^n\right], \label{power-law-profile}
\end{align}
where $n$ is a natural number. 
$r_*$ measures the length scale of inhomogeneity of the density distribution.
We introduce another length parameter $r_1$, the radius of a density region higher than the surrounding Friedmann-Lema\^itre-Robertson-Walker  background. 
$r_1$ and $r_*$ are independent.

\begin{figure}[htbp]
  \begin{center}
    \begin{tabular}{c}

      % 1
      \begin{minipage}{0.5\hsize}
        \begin{center}
          \includegraphics[scale=0.5]{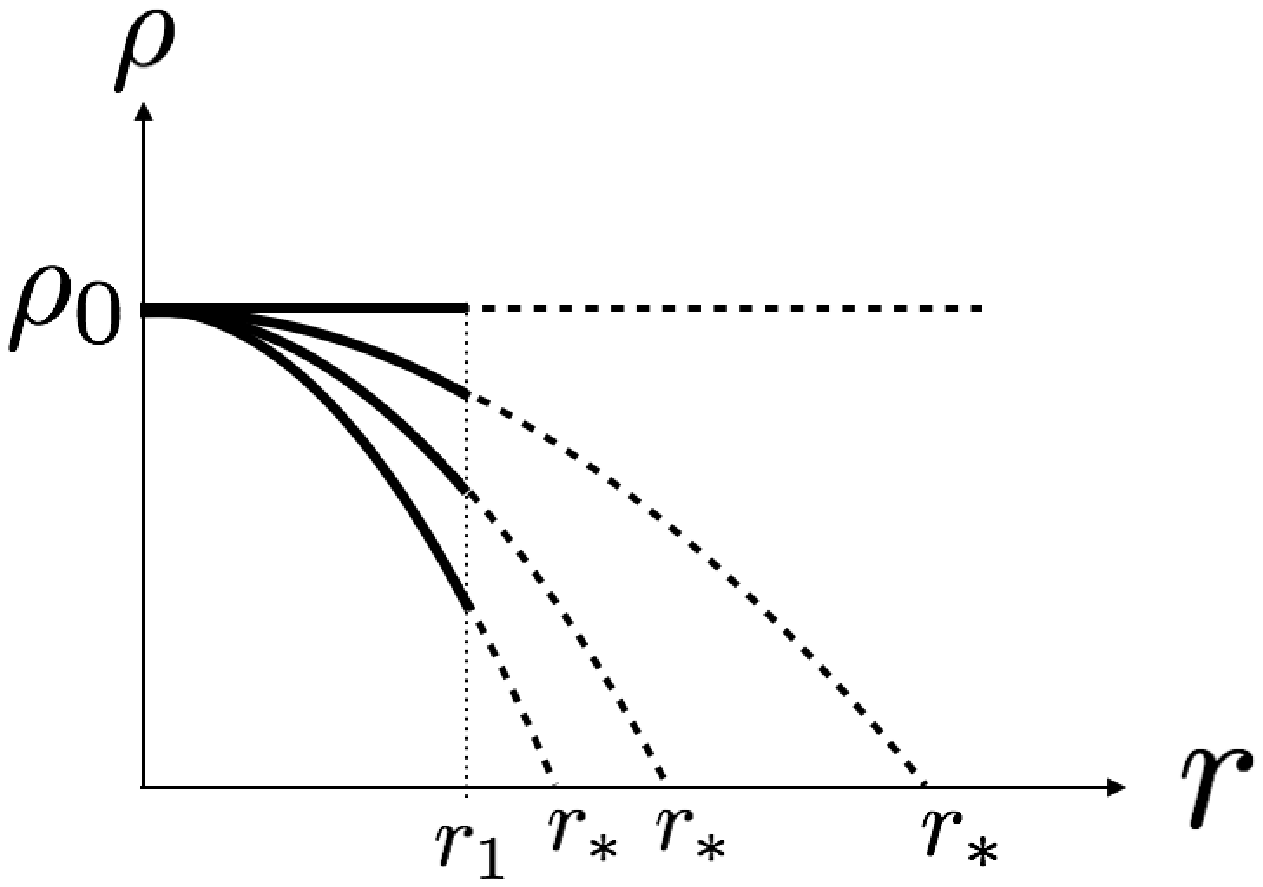}
          (a) 
        \end{center}
      \end{minipage}
      
      % 2
      \begin{minipage}{0.5\hsize}
        \begin{center}
          \includegraphics[scale=0.5]{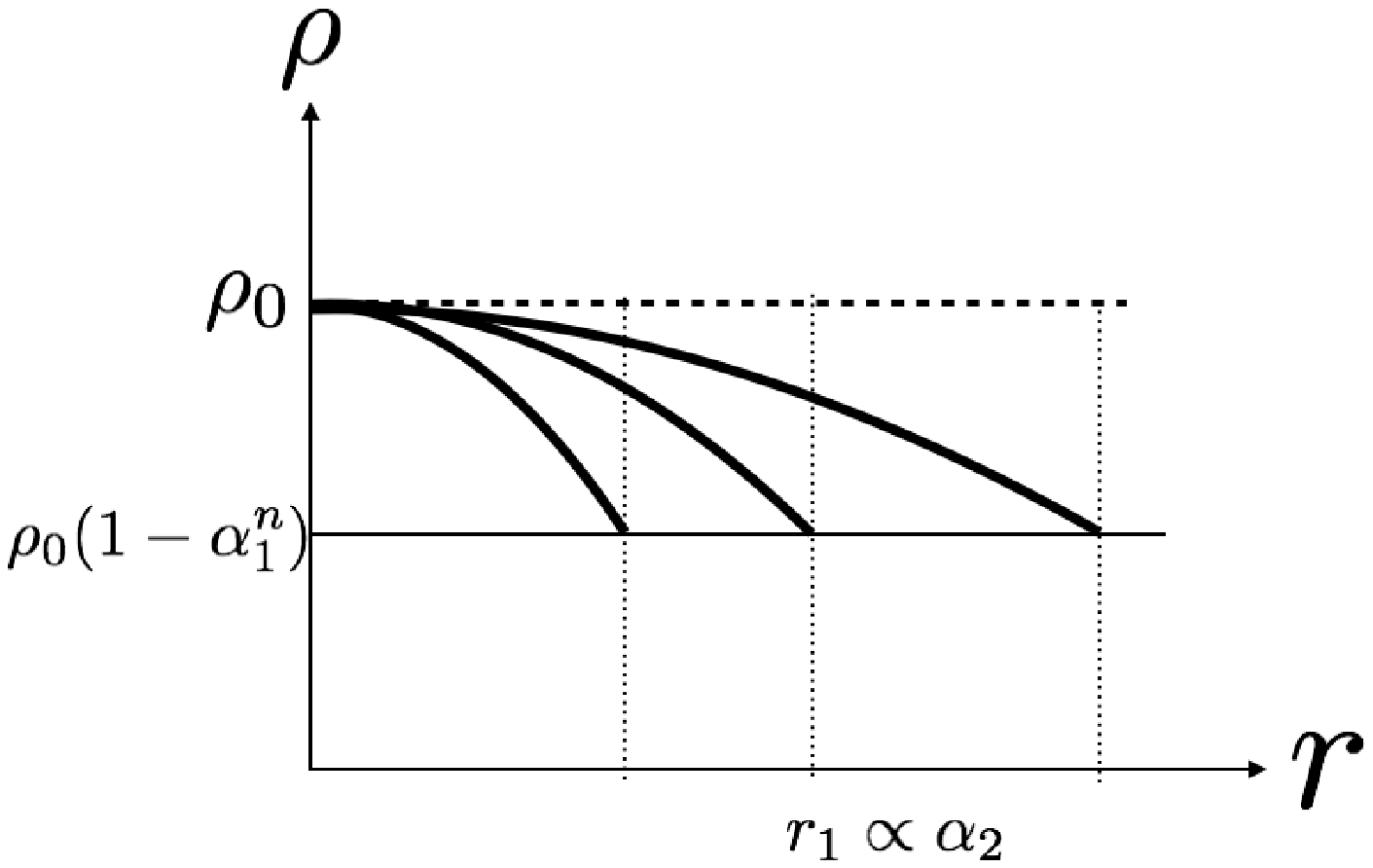}
           (b) 
        \end{center}
      \end{minipage}
 
\\
    \end{tabular}
    \caption{(a) The model of configurations (solid curves) represented by \eq{power-law-profile}. The model applies up to the radius $r_1$, a parameter determining the radius of high density region of the configuration. 
 $r_*$ measures the length scale of the density distribution.
(b) The configuration model for a fixed $\alpha_1$. A small value of $\alpha_2$ implies a configuration with a small radius.}
\label{fig-configuration}
  \end{center}
\end{figure}

The profile of \eq{power-law-profile} applies up to the radius $r_1$.
For $r_1/r_*=0$, this distribution is homogeneous, while it is highly inhomogeneous for $r_1/r_*=1$. 
Figure~\ref{fig-configuration} (a) explains $r_1, r_*$, and $\rho_0$ in a schematic manner.
It is known that black hole forms for $n\ge 4$ irrespective of $\rho_0$ and $r_*$ \cite{JoshiDwivedi1993}.
For $n=1$ and 2, the central singularity is at least locally naked, i.e., there are outgoing null geodesics that terminate in the past at the central singularity. Nakedness depends on $\rho_0$ and $r_*$ for  $n=3$
(see Ref. \cite{JoshiDwivedi1993, Barve+1999, Joshi1987book} for the details).  

%%%%%%%%%%%

We define that a configuration becomes a naked singularity if a null geodesic emanating from the center of the configuration reaches $r=r_1$ ~
\footnote{After reaching the radius $r=r_1$, the null geodesic trivially escapes to infinity ($r = \infty$) by propagating on the Friedmann-Lema\^itre-Robertson-Walker spacetime.}.
As mentioned in introduction, we should consider the time that a null geodesic takes to propagate from the center to the  surface of the configuration, i.e., we need to deal with a null geodesic propagating between two characteristic radii, $r=0$ and $r=r_1$. 
By introducing the equation of the null geodesic as $t=t_{null}(r)$ with $t_{null}(0)=t_s(0)$, we refine \eq{PK-condition} as
\begin{align}
t_{ah}(r_1)<t_{null}(r_1). \label{PBH-condition-revised}
\end{align}

\subsection{Equation of radial null geodesics in LTB spacetime}
From \eq{LTB-metric}, outgoing radial null geodesics  are found to obey the equation
\begin{align}
\frac{\D t_{null}(r)}{\D r}=\frac{R'(t_{null}(r),r)}{(1+f(r))^{1/2}} \label{outgoing-null}
\end{align}
As we consider momentarily static cases and have set $R(0,r)\equiv r$, by use of \eq{frRg}, \eq{outgoing-null} reduces to
\begin{align}
\frac{\D t_{null}(r)}{\D r}=\frac{R'(t_{null}(r),r)}{(1-R_g(r)/r)^{1/2}}. \label{outgoing-null-bound}
\end{align}
On the other hand, the total derivative of $R(t, r)$ is obtained as 
\begin{align}
\D {R(t,r)}=\dot{R}(t,r)\D t+R'(t,r)\D r.
\end{align}
$\dot{R}$ and $R'$ are given by differentiation of \eq{EOM-bound}. 
Since we are interested in $R(t_{null}(r),r)$, which corresponds to the areal radius evaluated on the outgoing null geodesic, we can write
\begin{align}
 \frac{\D  R(t_{null}(r),r)}{\D r}=\dot{R}(t_{null}(r),r)\frac{\D t_{null}(r)}{\D r}+R'(t_{null}(r),r).
\label{null-R}
\end{align}

\subsection{Null geodesics from $r=r_1$ to $r=0$}
To derive a criterion of PBH formation, we numerically investigate the gravitational collapse.
For convenience, we define dimensionless variables and parameters as follows:
\begin{align}
\tilde r:=\frac{r}{r_1}, \quad
\tilde R(\tilde{t},\tilde{r}):=\frac{R (\tilde{\rho}_{0}^{-1/2}\tilde{t},r_{1}\tilde{r})}{r_1}, \quad
\tilde t:=\tilde \rho_0^{\frac{1}{2}} t, \quad
\alpha_1:=\frac{r_1}{r_*}  \quad
\alpha_2 :=\tilde \rho_0^{\frac{1}{2}} r_1,
\end{align}
where $\tilde \rho_0$ is a scaled initial central density:
\begin{align}
\tilde \rho_0:=\frac{8\pi}{3}\rho_0.
\end{align}
The parameter $\alpha_1$ takes $0\leq \alpha_1 \leq 1$ and characterizes the degree of inhomogeneity. When $\alpha_1=0$ the configuration is completely homogeneous, while $\alpha_1 \simeq 1$ indicates a highly inhomogeneous configuration.

The gravitational radius $\tilde R_g(\tilde r)$, given in a dimensional form in  \eq{mass},  under the density profile \eq{power-law-profile} is written by
\begin{align}
&\tilde R_g (\tilde r):=\frac{R_g(r_1 \tilde r)}{r_1}=\alpha_2^2 \tilde r^3 A(\tilde r), \nonumber \\
&A(\tilde r):=1-\frac{3}{n+3}\alpha_1^n \tilde r^n.
\end{align}
The compactness $x$ and inhomogeneity $u$ are then written as
\begin{align}
x&=\alpha_2^2 A(1), \label{x}\\
u&=\frac{3}{n+3}\alpha_1^n A(1)^{-1}. \label{u}
\end{align}
$\alpha_1$ measures the inhomogeneity of the configuration, and $u \propto \alpha_1^n$ when $\alpha_1 \ll 1$. $\alpha_2$ measures the compactness of the high density region of the configuration (see \fig{fig-configuration} (b)).

The equation of motion of each shell, trajectory of the singularity $t_s$ and trajectory of the apparent horizon $t_{ah}$ can be reduced to a dimensionless form as
\begin{align}
&\tilde t=\tilde t_s(\tilde r)-A(\tilde r)^{-1/2}\left(\frac{\tilde R(\tilde t, \tilde r)}{\tilde r}\right)^{3/2}F\left(\frac{\tilde R(\tilde t, \tilde r)}{\tilde r}\right), \label{tilde-t} \\
&\tilde t_s(\tilde r)=\frac{\pi}{2}A(\tilde r)^{-1/2},\\
&\tilde t_{ah}(\tilde r)=\tilde t_s(\tilde r)-(\alpha_2\tilde r)^3A(\tilde r) F\left(\alpha_2^2\tilde r^2A(\tilde r)\right). 
\label{tah-tilde}
\end{align}

\begin{figure}[t]
\begin{center}
\includegraphics[clip,width=8cm]{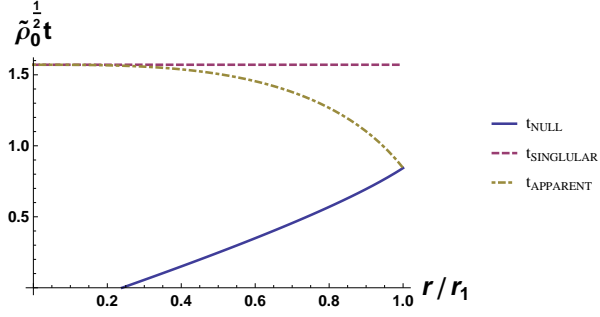}
\caption{Null geodesic, the trajectory of the singularity and the trajectory of the apparent horizon of the Oppenheimer-Snyder collapse. All the point inside $r=r_1$ becomes singular at $t=t_s=\pi/2$.}
\label{fig-t_OS} 
\end{center}
\end{figure}

As a simple example, we take the gravitational collapse of a homogeneous configuration and show how $\tilde t_s, \tilde t$ and $\tilde t_{ah}$ behave. This collapse is known as the Oppenheimer-Snyder collapse \cite{Oppenheimer-Snyder} and  is described by taking $\alpha_1=0$. In this case $t=t_s(r)$ and $t=t_{ah}(r)$ behave typically like \fig{fig-t_OS}.
From the figure, one can see that the  surface $r=r_1$ is trapped first and the smaller radius is trapped later. The center is trapped last. The center becomes singular after the surface $r=r_1$ becomes trapped. 
Indeed, one finds from \eq{tah-tilde} that $\tilde t_{ah}(0)=\tilde t_s(0)$.
Hence the collapse with complete homogeneity, the Oppenheimer-Snyder model, leads to a black hole formation.

We integrate the ordinary differential equation for $R_{null}$, \eq{null-R}, and substitute the solution into \eq{tilde-t} to obtain the null geodesic.
We compute $\partial \tilde R/\partial \tilde r$ and $\partial  R/\partial  t$ by differentiating \eq{tilde-t} as
\begin{align}
\frac{\partial \tilde R}{\partial \tilde r}&
=\frac{3n}{2(n+3)}\frac{\alpha_1^n{\tilde r}^n}{A}\left(\frac{\tilde r}{\tilde R}-1 \right)^{1/2}\left[\frac{\pi}{2}-\left(\frac{\tilde R}{\tilde r}\right)^{3/2}F\left(\frac{\tilde R}{\tilde r}\right) \right]+\frac{\tilde R}{\tilde r}, \\
\frac{\partial  \tilde R}{\partial \tilde t}&
=-\frac{1}{\alpha_2}\sqrt{\frac{\tilde R_g}{\tilde R}}\sqrt{1-\frac{\tilde R}{\tilde r}}.
\end{align}
Then we derive the differential equation for $\tilde{R}(\tilde{t}_{null}(\tilde{r}),\tilde{r})$, the areal radius along null geodesics, as
\begin{align}
 \frac{\D \tilde R(\tilde t_{null}(\tilde r),\tilde r)}{\D \tilde r}
=R'(\tilde{t}_{null}(\tilde{r}),\tilde r)\left[
1
-\sqrt{\frac{\tilde R_g(\tilde r)}{\tilde R(\tilde{t}_{null}(\tilde r), \tilde r)}}
\sqrt{\frac{1-\tilde R(\tilde{t}_{null}(\tilde r),\tilde r)/\tilde r}{1-\tilde R_g(\tilde r)/\tilde r}} \right]
\label{null-R-substituted}
\end{align}
Because \eq{null-R-substituted} is in general not analytically solvable, we integrate it numerically. 
In order to investigate whether information from the central singularity reaches the surface $r=r_1$ at a time when the surface is covered by the apparent horizon, we emit lights from the surface $r=r_1$ to the center $r=0$ in the past direction
\footnote{Although one can emit lights from the center to the surface in the future direction like in Ref. \cite{JoshiDwivedi1993}, it takes additional amount of labor in calculation.}.
Thus, the boundary condition we impose is 
\begin{align}
R(t_{null}(r_{1}),r_1)=R_g(r_1) \Leftrightarrow \tilde R(\tilde{t}_{null}(1),1)=\tilde R_g(1). 
\end{align}

\begin{figure}[t]
\begin{center}
\includegraphics[clip,width=11cm]{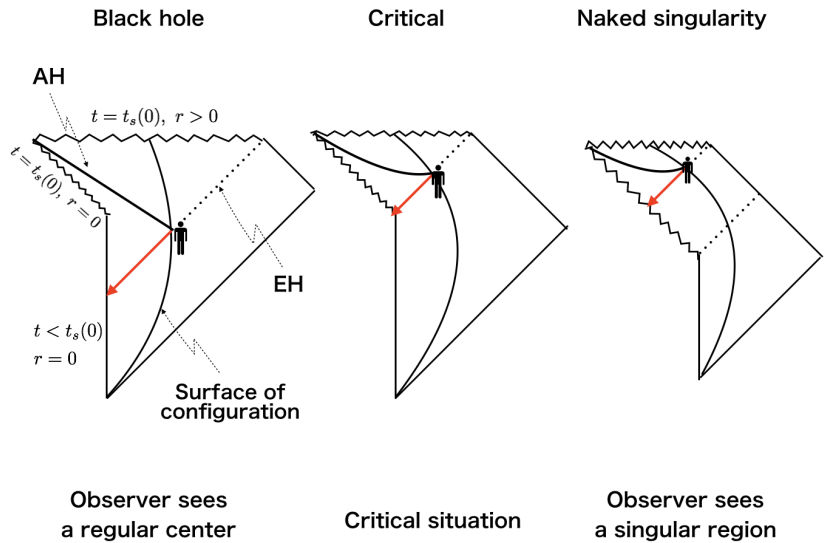}
\caption{Penrose diagram of the black hole formation (left), the naked singularity formation (right) and the critical situation (middle). Solid thick curves and dotted straight lines are apparent horizons and event horizons, respectively. Red arrows represent past directed null geodesics starting from the surfaces of configurations to the center.
}
\label{fig-BH-Naked} 
\end{center}
\end{figure}

The critical situation separates the collapse scenario into two cases, that is, the black hole formation and the naked singularity formation.
Figure \ref{fig-BH-Naked} depicts the difference among the black hole formation, the naked singularity formation and the critical situation
\footnote{These Penrose diagrams describe asymptotically flat universe, although our real universe is not asymptotically flat. However, since we are interested in a local phenomenon, these pictures are sufficient for our explanation. }.
For the black hole formation, an observer fixed at the surface of the configuration sees a regular center. For the naked singularity formation, the observer would see the central singularity. The critical situation is the case that the observer would marginally see the central singularity, which separates the fate of a collapse into black hole formation and singularity formation.

\begin{figure}[t]
  \begin{center}
    \begin{tabular}{c}

      % 1
      \begin{minipage}{0.33\hsize}
        \begin{center}
          \includegraphics[scale=0.66]{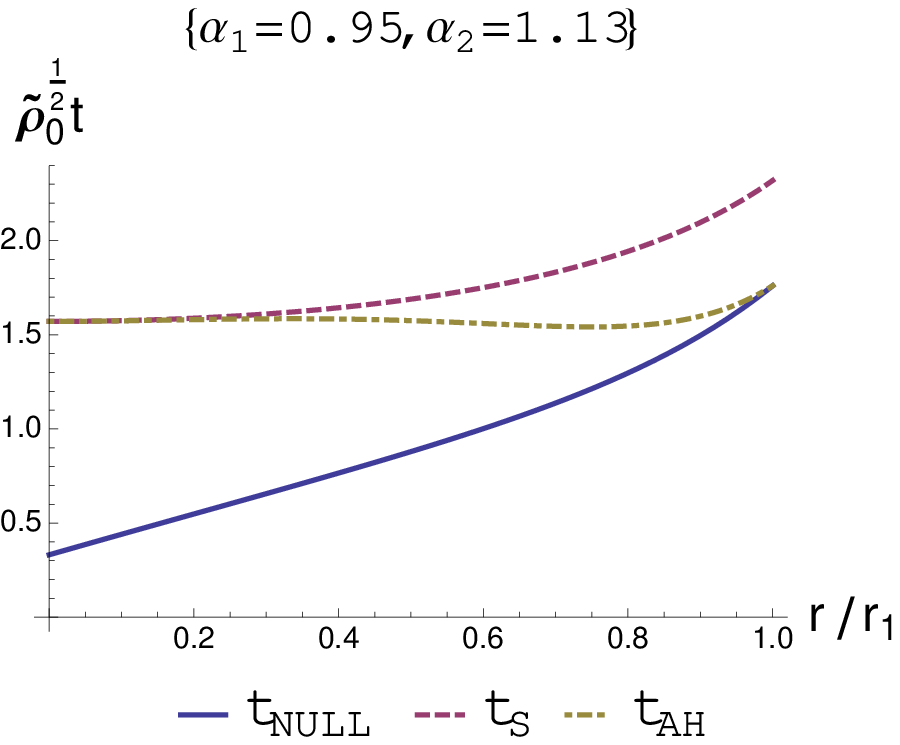}
           (a) 
        \end{center}
      \end{minipage}

      % 2
      \begin{minipage}{0.33\hsize}
        \begin{center}
          \includegraphics[scale=0.45]{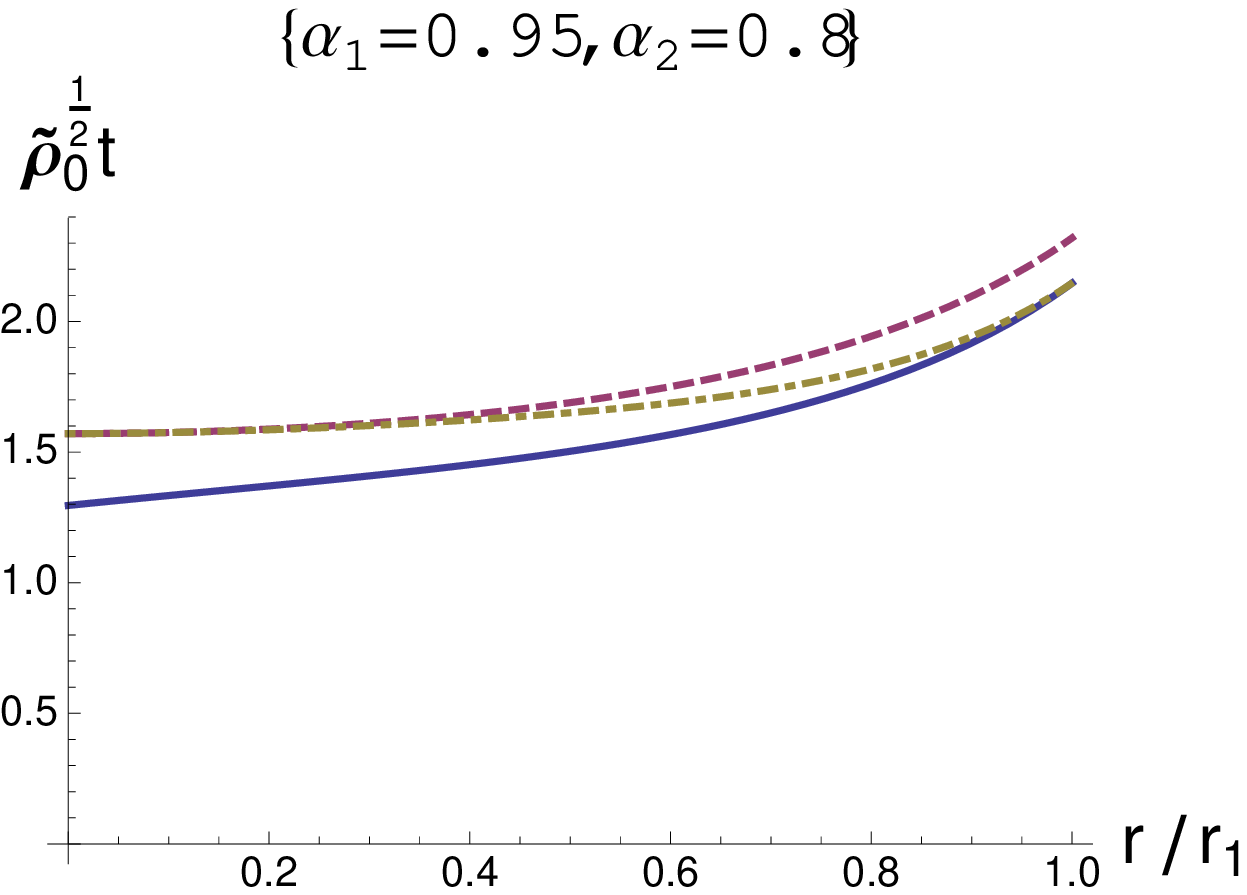}
           (b) 
        \end{center}
      \end{minipage}
      
      % 3
      \begin{minipage}{0.33\hsize}
        \begin{center}
          \includegraphics[scale=0.45]{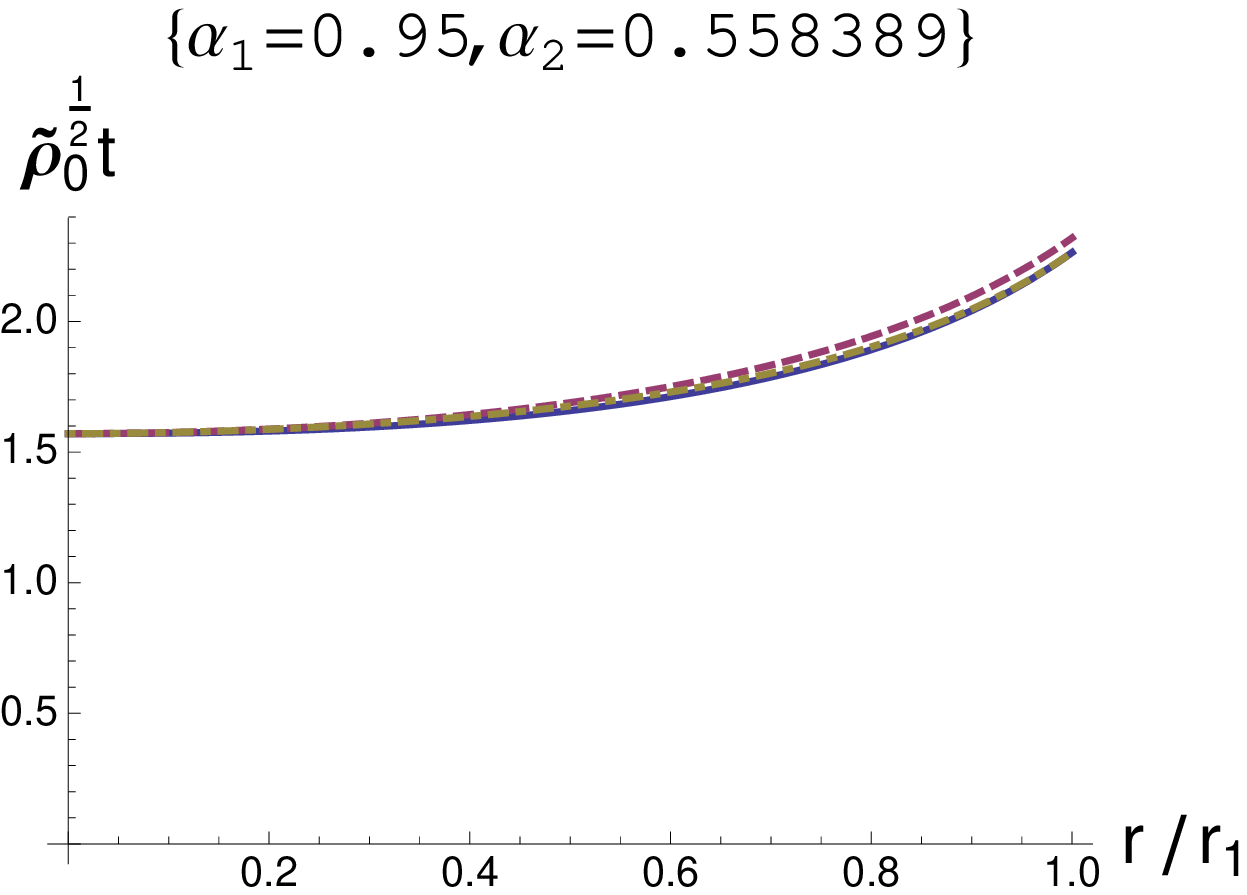}
          (c) 
        \end{center}
      \end{minipage}

\\
    \end{tabular}
    \caption{Typical behavior of null geodesics with a typical parameter, $\alpha_1=0.95$.
    Solid curves are the past-directed ingoing null geodesic starting from the surface $r=r_1$ at
   $t=t_{ah}(r_{1})$, or equivalently, the future-directed outgoing null propagating from the center to reach the surface $r=r_1$ at $t=t_{ah}(r_{1})$. 
   Dashed curves and dot-dashed curves represent the trajectory of the singularity and that of apparent horizon, respectively. 
(a): The null geodesic from $r=r_1$ reaches the regular center $r=0$. This indicates a black hole formation. 
(b): In the same way as the case (a), null geodesic reaches the regular center and hence this is also a black hole formation. However, the difference between $\tilde t_s(0)$ and $\tilde t_{null}(0)$ decreases as the value of $\alpha_2$ decreases. 
(c):  The null geodesic hits the singularity at $\alpha_2= 0.558389$. 
Hence the critical parameter is found to be $\alpha_2^{crit}= 0.558389$ for $\alpha_1=0.95$.
The case (a) and (b) were not black hole formation if the condition of Ref. \cite{PolnarevKhlopov1981} was adopted.}    
\label{fig-null-t-1}
  \end{center}
\end{figure}

Whether PBH forms or not depends on the set of the two parameters $(\alpha_1, \alpha_2)$.
We search the sets of $\left(\alpha_1, \alpha_2^{crit}(\alpha_1)\right)$ where $\alpha_{2}^{crit}(\alpha_1)$ corresponds to the critical situation for a given value of $\alpha_1$.
The critical set of parameters $\left(\alpha_1, \alpha_2^{crit}(\alpha_1)\right)$ can be translated to $\left(x(\alpha_1, \alpha_2^{crit}(\alpha_1)), u(\alpha_1)\right)$ via \eq{x} and \eq{u}.
We show typical behaviors of null geodesics with typical parameters in \fig{fig-null-t-1}.
A physical interpretation on typical behaviors of null geodesics is following; \fig{fig-configuration} (b) shows that smaller values of $\alpha_2$ indicate the smaller size of configurations for fixed $\alpha_1$ and $\rho_0$. Since null geodesics emanating from the center easily escape to infinity when the size of configuration is small, small values of $\alpha_2$ help to form naked singularities.

%%%%%%%%%%%
%%%%%%%%%%%

\subsection{Result}
\begin{figure}[htbp]
  \begin{center}
    \begin{tabular}{c}

      % 1
      \begin{minipage}{0.5\hsize}
        \begin{center}
          \includegraphics[scale=0.7]{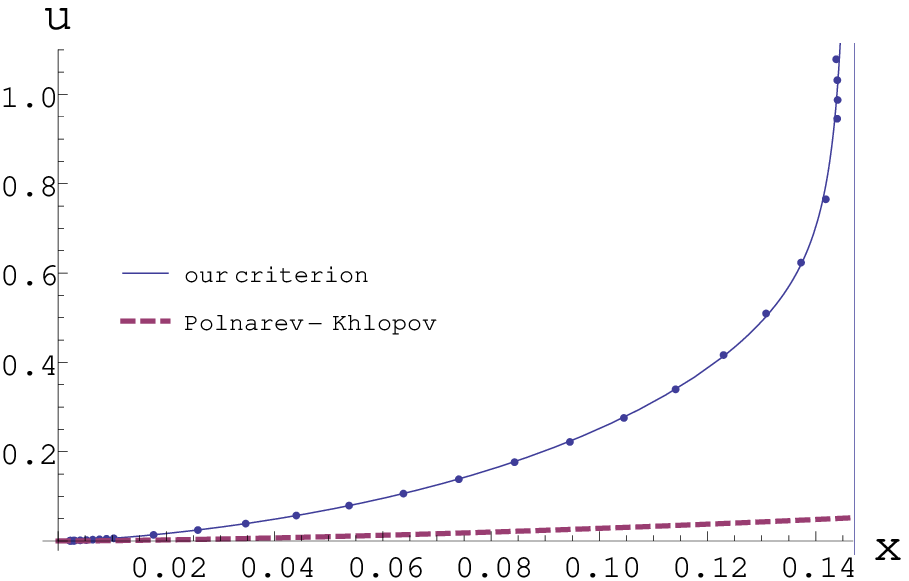}
          (a) 
        \end{center}
      \end{minipage}
      
      % 2
      \begin{minipage}{0.5\hsize}
        \begin{center}
          \includegraphics[scale=0.5]{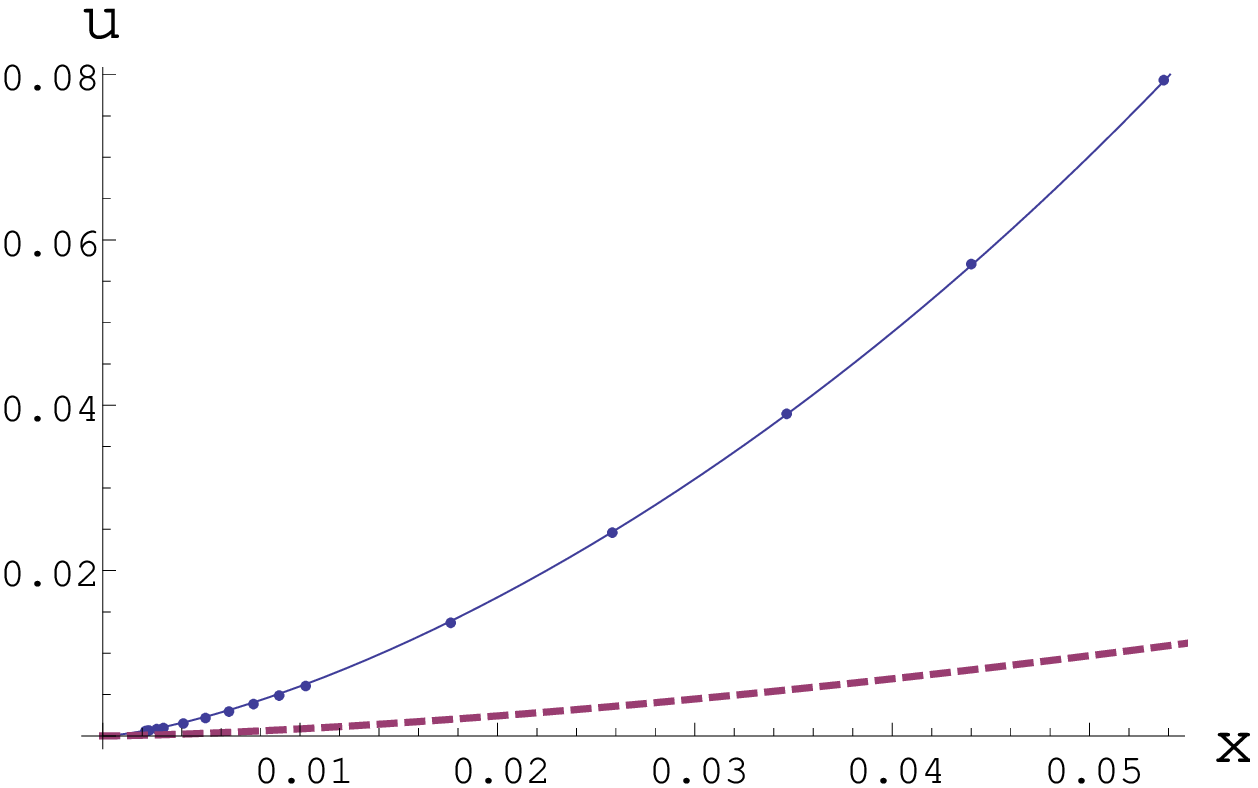}
         (b) 
        \end{center}
      \end{minipage}
 
\\
    \end{tabular}
    \caption{(a) Critical points $\left(x(\alpha_1, \alpha_2^{crit}(\alpha_1)), u(\alpha_1)\right)$ in the range $0<x< x_c$. The solid curve is an interpolating function. 
The dashed curve is given by \eq{PK-condition-variant-variant} which is derived in Ref. \cite{KhlopovPolnarev1980}.
  (b) Close-up of the figure (a) around the origin. The solid curve is given by \eq{leading-function} where the power-law index is derived by the previous study \cite{KhlopovPolnarev1980}  without identifying the proportionality coefficient. It is also derived in this study as \eq{PK-condition-variant-variant}, shown by dashed curves.
}    
\label{fig-data}
  \end{center}
\end{figure}

We show the critical set of parameters $\left(\alpha_1, \alpha_2^{crit}(\alpha_1)\right)$, which is subsequently used to improve the PBH formation criterion.
For numerical calculation under the profile \eq{power-law-profile}, we took $n=2$ since this power is generic if we assume that the density field is regular.
In \fig{fig-data} (a), we plot the critical points  $\left(x(\alpha_1, \alpha_2^{crit}(\alpha_1)), u(\alpha_1)\right)$ in $(x, u)$ plane.
These points correspond to the critical situation in \fig{fig-BH-Naked}. 
For the range $0<x< x_c\simeq 0.147$, we  found that our numerical result can be fitted well by a rational function $f_2(x)$ drawn in \fig{fig-data} (a) as
\begin{align}
f_2(x):=
 \begin{cases}
    \frac{a_1 x^{3/2}+a_2 x^2}{b_1+b_2 x+b_3 x^{3/2}} & (0<x< x_c) \\
    \infty & (x_c \leq x)
  \end{cases}
, \label{fitting-function}
\end{align}
where $a_1=3.1981, a_2=-8.2603, b_1=0.4569, b_2=-7.2461, b_3=10.797$. 
Thus, the criterion of PBH formation is given by
\begin{align}
u < f_2(x).
\label{improved-criterion}
\end{align}
For $x\ll 1$, \eq{fitting-function} reduces to
\begin{align}
f_2(x) \simeq 7.0003x^{3/2}. \label{leading-function}
\end{align}
to the leading order of $x$. \eq{leading-function} is plotted in \fig{fig-data} (b).
All the region of $(x,u)$ below the interpolating function results in PBH formation, while the region above it results in a naked singularity formation. 
We found that the interpolating function is monotonically increasing and diverges at $x=x_c$.  Beyond $x_c$ there is no critical point and this means any configuration forms a black hole.

%%%%%%%%%%%

From the agreement of  \eq{leading-function} with numerical results depicted in \fig{fig-data}(b), we found that the relation $f_2(x) \propto x^{3/2}$ derived by Khlopov-Polnarev agrees well with our numerical result for sufficiently small $x$, while the normalization depends on the proportionality coefficient. 
Let us compare \eq{leading-function} with \eq{PK-condition-approx}. The coefficients of the two functions are 0.85 and 7.00, respectively. 
This difference is caused by the finite speed of a signal propagating from the central region to the surface of a configuration.
 The signal speed is taken into account in \eq{leading-function}, while it is not in \eq{PK-condition-approx}.

\section{PBH Production Probability}\label{PBH Production Probability}
In this section we evaluate the probability of PBH formation as a function of the standard deviation $\sigma$ of cosmological density perturbation.
We denote the production probability derived by considering only the effect of inhomogeneity as $\beta_{inhom}$.
We mainly focus on $\beta_{inhom}$ and comment on the probability including $\beta_{aniso}$, the probability associated with the effect of anisotropy of collapsing matters, in the last part of this section.

\begin{figure}[htb]
\begin{center}
\includegraphics[clip,width=10cm]{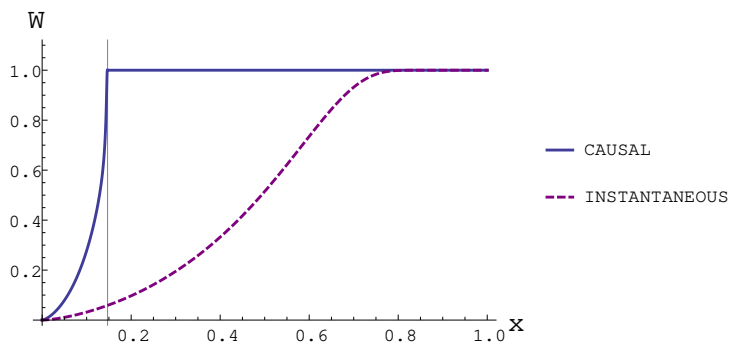}
\caption{Probability of PBH formation as a function of the compactness. The solid curve is our result of \eq{W-with-signal}, while the dashed curve is given by \eq{W-without-signal}. $W^{caus}$ becomes unity at $x=x_c$.}
\label{fig-Probability} 
\end{center}
\end{figure}

We evaluate $\beta_{inhom}$ and compare the probability derived with our criterion of PBH formation with that derived by the criterion of Ref. \cite{PolnarevKhlopov1981}.
To compare them, it is convenient to use $W(x)$, a production probability in terms of $x$.
We assume the Gaussian distribution restricted to $u \ge 0$ for inhomogeneity following previous researches \cite{KhlopovPolnarev1980, PolnarevKhlopov1981}. 
For our configuration model, \eq{power-law-profile}, $W(x)$ is given by
\begin{align}
W(x)&=\frac{2}{\sqrt{2\pi \Sigma^2}}\int^{u_{max}}_0 \D u\exp{\left[-\frac{u^2}{2\Sigma^2}\right]}
={\rm Erf}\left(\frac{u_{max}}{\sqrt{2} \Sigma}\right), \label{W}
\end{align}
where ${\rm Erf}(X)$ is the error function and $u_{max}$ is the maximum value of $u$ that allows PBHs to form.
In Ref. \cite{KhlopovPolnarev1980, PolnarevKhlopov1981}, $\Sigma\sim 1$ was assumed.
Since there is no affirmative reason to take $\Sigma\sim 1$, we leave $\Sigma$  as it is in the formulation.
We first evaluate $W^{inst}(x)$ in which the signal propagation is effectively assumed to be instantaneous.
By substituting \eq{PK-condition-variant-variant} into \eq{W}, we obtain 
\begin{align}
W^{inst}(x)=
\begin{cases}
    {\rm Erf}\left[f_1(x)/(\sqrt{2} \Sigma)\right] & (0<x<1) \\
    1 & (1\leq x)
  \end{cases}
\label{W-without-signal}
\end{align}
Since any configuration with $x\geq 1$ represents a black hole at the initial state, $W^{inst}(x\geq1)=1$ is trivial.
We next evaluate $W^{caus}(x)$ in which the signal speed is taken into account, i.e., causality is considered.
In this case, by using \eq{fitting-function} we obtain 
\begin{align}
W^{caus}(x) =
\begin{cases}
     {\rm Erf}\left[ f_2(x)/(\sqrt{2} \Sigma) \right] & (0<x< x_c) \\
    1 & (x \geq x_c\simeq 0.147)
  \end{cases}
   .\label{W-with-signal}
\end{align}
\eq{W-without-signal} and \eq{W-with-signal} are plotted in \fig{fig-Probability}.
This figure shows that $W^{caus}$ goes to unity at smaller values of $x$ than $W^{inst}$ does.
Only in carrying out calculations we assume $\Sigma=1$ following Ref. \cite{KhlopovPolnarev1980}.

\subsection{PBH production probability with inhomogeneity}
To derive a production probability $\beta_{inhom}$ of PBHs, we must specify a relation between the parameter $x$ and $\delta_H$, the value of cosmological density perturbation when the configuration enters the Hubble horizon. 
It is shown in Ref. \cite{HaradaYooKohriNakaoJhingan2016} that (see their Eq.(17))
\begin{align}
x\approx \frac{4}{3}\delta_H \label{x-delta-relation}
\end{align}
for $x\ll 1$.
On the other hand, the PBH production probability $\beta_{inhom}(\sigma)$ from inhomogeneous configurations, where $\sigma$ is the standard deviation of $\delta_H$, is obtained by using \eq{W} as
\begin{align}
\beta_{inhom}(\sigma)
&:=\frac{1}{\sqrt{2\pi \sigma^2}}\int^{\infty}_0\D \delta_H \exp{\left[-\frac{\delta_H^2}{2\sigma^2}\right]}
 W\left(x(\delta_H)\right). \label{BETA-inhom}
\end{align}
Here, we first derive the production probability $\beta_{inhom}^{inst}(\sigma)$ by assuming instantaneous propagation of information by substituting \eq{W-without-signal} into \eq{BETA-inhom} as

\begin{align}
\beta_{inhom}^{inst}(\sigma)
&:=\frac{1}{\sqrt{2\pi \sigma^2}}\int^{\infty}_0\D \delta_H \exp{\left[-\frac{\delta_H^2}{2\sigma^2}\right]}
 W^{inst}\left(x(\delta_H)\right). \label{beta-inhom-wo}
\end{align}
In the derivation of $\beta_{inhom}^{inst}$, we simply applied \eq{x-delta-relation} for $x \sim 1$, although there is no guarantee that such an extrapolation is justified. 
Next, we derive the production probability including causal propagation of information by substituting \eq{W-with-signal} into \eq{BETA-inhom} as
\begin{align}
\beta_{inhom}^{caus}(\sigma)
&:=\frac{1}{\sqrt{2\pi \sigma^2}}\int^{\infty}_0\D \delta_H \exp{\left[-\frac{\delta_H^2}{2\sigma^2}\right]}
 W^{caus}\left(x(\delta_H)\right) \nonumber \\
&=\frac{1}{\sqrt{2\pi \sigma^2}}\int^{\delta_H(x_c)}_0\D \delta_H \exp{\left[-\frac{\delta_H^2}{2\sigma^2}\right]}
 W^{caus}\left(x(\delta_H)\right)
+ \frac{1}{2}\left[1-{\rm Erf}\left(\frac{\delta_H(x_c)}{\sqrt{2}\sigma}\right)\right], \label{beta-inhomo}
\end{align}
where $\delta_H(x_c) \approx 0.11$. 
In the derivation of $\beta_{inhom}^{caus}$, unlike the derivation of $\beta_{inhom}^{inst}$, we may safely use \eq{x-delta-relation} because $x_c\simeq 0.1$ is not so large compared to unity.

\begin{figure}[t]
\begin{center}
\includegraphics[clip,width=7cm]{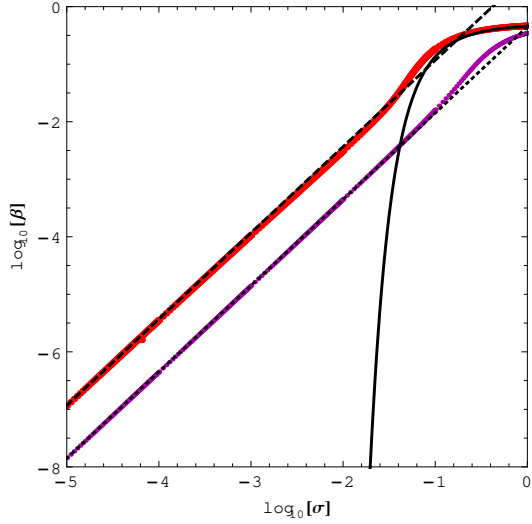}
\caption{Production probability $\beta_{inhom}(\sigma)$ of PBH by collapsing inhomogeneous density distributions in the matter dominated era, where $\sigma$ is the standard deviation of density distributions. 
Note that this probability includes only the effect of inhomogeneity. The effects of anisotropy and spin are not included.
The purple curve  represents the probability $\beta_{inhom}^{inst}$ which does not include the effect of the signal speed, while the red curve is $\beta_{inhom}^{caus}$ in which the signal speed is  considered.
We found both  $\beta_{inhom}^{caus}$ and $\beta_{inhom}^{inst}$ are monotonically increasing functions of $\sigma$.
The figure shows that $\beta_{inhom}^{inst}<\beta_{inhom}^{caus}$.
For $\sigma \ll 1$, $\beta_{inhom}^{caus}$ and $\beta_{inhom}^{inst}$ can be approximated by $3.6979\sigma^{3/2}$ (dashed black line) and $0.4484\sigma^{3/2}$ (dotted black line), respectively. 
Their powers of $3/2$ was originally derived by Khlopov and Polnarev. 
For $\sigma \sim \mathcal O(1)$, $\beta_{inhom}^{caus}$ is approximated by the error function (black curve).
}
\label{fig-BETA} 
\end{center}
\end{figure}

We plot $\beta_{inhom}^{inst}(\sigma)$ and $\beta_{inhom}^{caus}(\sigma)$  in \fig{fig-BETA} with the assumption $\Sigma=1$. 
As seen from the figure, both  $\beta_{inhom}^{caus}$ and $\beta_{inhom}^{inst}$ are monotonically increasing functions of $\sigma$.
The figure also shows that $\beta_{inhom}^{inst}<\beta_{inhom}^{caus}$.
For $\sigma \ll 1$, $\beta_{inhom}^{caus}$ is larger by about an order of magnitude than $\beta_{inhom}^{inst}$. Each production probability seems to be approximated by a power law with a common index (see below). 
For $\sigma \sim \mathcal O(1)$, the difference between $\beta_{inhom}^{inst}(\sigma)$ and $\beta_{inhom}^{caus}(\sigma)$ decreases, because all the configuration forms a black hole.

\subsubsection{Semi-analytic formula}
We derive a semi-analytic formula of $\beta_{inhom}^{caus}$ for $\sigma \ll 1$.
In this case, the second term of \eq{beta-inhomo} vanishes because $\delta_H(x_c)/\sigma \gg 1$.
Only the integration range $0<\delta_H\ll 1$ contributes substantially to the first term.
 Then, with the use of \eq{leading-function}, $\beta_{inhom}^{caus}$ reduces to
\begin{align}
\beta_{inhom}^{caus}(\sigma)
&\simeq \frac{1}{\sqrt{2\pi \sigma^2}}\int^{\delta_H(x_c)}_0\D \delta_H \exp{\left[-\frac{\delta_H^2}{2\sigma^2}\right]} 
\times  \frac{56}{3\Sigma}\sqrt{\frac{2}{3 \pi }}\delta_H ^{3/2}
=\frac{56}{3\sqrt{3} \pi \Sigma} \sigma^{3/2} \int^{\delta_H(x_c)/\sigma}_0 \D y \exp{\left[-\frac{y^2}{2}\right]}  y^{3/2}  \nonumber \\ 
&\simeq \frac{56}{3\sqrt{3} \pi \Sigma}2^{1/4}\Gamma \left(\frac{5}{4}\right)  \sigma^{3/2}
=3.6979 \frac{\sigma^{3/2}}{\Sigma}  \qquad \qquad  (\sigma \ll 1),
 \label{small-beta}
\end{align}
where $y:=\delta_H/\sigma$ and $\Gamma(X)$ is the gamma function. 
In the same way, $\beta_{inhom}^{inst}(\sigma)$ for $\sigma \ll 1$ is obtained as 
\begin{align}
\beta_{inhom}^{inst}(\sigma)\simeq 0.4484 \frac{\sigma^{3/2}}{\Sigma} \qquad \qquad  (\sigma \ll 1).
\label{small-beta-without}
\end{align}
The power $3/2$ in \eq{small-beta} and \eq{small-beta-without} was derived by Khlopov and Polnarev \cite{KhlopovPolnarev1980, PolnarevKhlopov1981}. 
The approximated function of \eq{small-beta} is valid up to $\sigma \sim 0.05$, while \eq{small-beta-without} is valid up to $\sigma \sim 0.1$.

We also derive an analytic formula of $\beta_{inhom}^{caus}$ for $\sigma \sim \mathcal O(1)$. In this case the second term of \eq{beta-inhomo} becomes dominant, i.e., 
\begin{align}
\beta_{inhom}^{caus}(\sigma)
\simeq \frac{1}{2}\left[1-{\rm Erf}\left(\frac{0.11}{\sqrt{2}\sigma}\right)\right]
\qquad \qquad  (\sigma \sim \mathcal O(1) ).
\label{large-beta}
\end{align}
We plot \eq{small-beta}, (\ref{small-beta-without}) and (\ref{large-beta}) with the assumption $\Sigma=1$ in \fig{fig-BETA}.

\subsection{Production probability including inhomogeneity and anisotropy}
\label{section-inhomo-aniso}
Here, we comment on the production probability including inhomogeneity and anisotropy effects.
We have calculated PBH production probability incorporating only the effect of inhomogeneity, $\beta_{inhom}$. 
The probability incorporating the effect of anisotropy of collapsing masses plays an important role for black hole formation.
Ref. \cite{HaradaYooKohriNakaoJhingan2016} investigated the effect of anisotropy for PBH formation. 
In Ref. \cite{HaradaYooKohriNakaoJhingan2016}, the production probability arising from the effect of anisotropy is given by $\beta_{aniso}(\sigma)$ as
\begin{align}
\beta_{aniso}(\sigma):=\int^{\infty}_{0} \D \alpha \int^{\alpha}_{-\infty} \D \beta \int^{\beta}_{-\infty} \D \gamma \left(1-h(\alpha, \beta, \gamma) \right)w(\alpha, \beta, \gamma)
\end{align}
with
\begin{align}
w(\alpha, \beta, \gamma)\D \alpha\D \beta \D \gamma=&-\frac{27}{8\sqrt{5}\sigma_3^6}\exp\left[-\frac{3}{5\sigma_3^2}\left\{ (\alpha^2+\beta^2+\gamma^2)-\frac{1}{2}(\alpha \beta+\beta \gamma+ \gamma \alpha) \right\}  \right]\nonumber \\
&\times 
(\alpha-\beta)(\beta-\gamma)(\gamma-\alpha)\D \alpha\D \beta \D \gamma, \\
h(\alpha, \beta, \gamma):=&\frac{2}{\pi}\frac{\alpha-\gamma}{\alpha^2}E\left(\sqrt{1-\left(\frac{\alpha-\beta}{\alpha-\gamma}\right)^2}\right),
\end{align}
where $\alpha, \beta$ and $\gamma$ characterize anisotropy of collapsing matters, $E(k)$ is the complete elliptic integral of the second kind and  $\sigma_3=\sigma/\sqrt{5}$~~
\footnote{For the peak statistics in the radiation dominated era, see~\cite{Yoo:2018kvb}.}. 

As written above, $\beta_{inhom}$ and $\beta_{aniso}$ have been calculated separately.
In general, the production probability $\beta_{inhom+aniso}$, which includes the effects of both inhomogeneity and anisotropy, depends on the four variables $(\alpha, \beta, \gamma, \delta_H)$ in a complicated manner.
However, if $\sigma$ is sufficiently small, we speculate that one can obtain $\beta_{inhom+aniso}$ in the following manner.
For $\sigma \ll 1$, $\beta_{inhom}$ is given by \eq{small-beta} and $\beta_{aniso}$ is semi-analytically given by \cite{HaradaYooKohriNakaoJhingan2016}
\begin{align}
\beta_{aniso}\simeq  0.05556 \sigma^5
\end{align}
Ref.~\cite{HaradaYooKohriNakaoJhingan2016} showed that most of the collapses which result in PBH formation must be nearly spherically symmetric.
In this case, we expect  $\beta_{inhom+aniso}$ may be given by the simple multiplication of $\beta_{inhom}$ and $\beta_{aniso}$,
\begin{align}
\beta_{inhom+aniso} \simeq \beta_{inhom}\times \beta_{aniso} =0.2055\sigma^{13/2}. \label{beta-total-small}
\end{align}

\color{black}
%%%%%%%%%%%%%%%%%%%%%%%%%%%%%%%%%%%%%%%%%
%%%SECTION%%%SECTION%%%SECTION%%%SECTION%%%SECTION%
%%%%%%%%%%%%%%%%%%%%%%%%%%%%%%%%%%%%%%%%%
\section{Summary and Discussions}\label{Conclusion-section}
Based on Khlopov and Polnarev's pioneering study, we refined the criterion of PBH formation in the matter dominated era. In their formulation, PBHs are formed if \eq{PK-condition-ux} holds.
As a first step to refinement of the criterion, we re-evaluated the condition of \eq{PK-condition} and then obtained \eq{PK-condition-variant-variant}.
Next, to derive the improved criterion of PBH formation, we included the effect of causal propagation of information from the central singularity to the surface of a density distribution. By considering this effect, we revised the criterion \eq{PK-condition-variant-variant} to \eq{improved-criterion}. We found that when the value $x$ characterizing the compactness of the configuration is small enough, the two criterions, \eq{PK-condition-variant-variant} and \eq{improved-criterion}, reduce to $u \lesssim 0.85x^{3/2}$ and $u \lesssim 7.00x^{3/2}$, respectively. 
The large difference of the two coefficients implies that signal/information propagation must be taken into account to evaluate black hole formation appropriately.

As a main result, we have derived the production probability $\beta_{inhom}$ as a function of $\sigma$, the standard deviation of cosmological density perturbation.
$\beta_{inhom}$ was calculated by applying the criterion of ``instantaneous" (\eq{PK-condition-variant-variant}) and of ``causal" (\eq{improved-criterion}) and we have compared $\beta_{inhom}^{inst}$ and $\beta_{inhom}^{caus}$ to identify how the consideration on the signal propagation affects the production probability.
We found both $\beta_{inhom}^{inst}$ and $\beta_{inhom}^{caus}$ are monotonically increasing functions of $\sigma$. 
We also found $\beta_{inhom}^{inst}<\beta_{inhom}^{caus}$, i.e., the consideration of the signal propagation increases the probability of black hole formation.
This increase is intuitively reasonable because our new condition can count overlooked cases in which an apparent horizon is formed before the signal from the central singularity reaches the surface of the configuration.

Our new probability $\beta_{inhom}^{caus}$ behaves as a power law of $3.6979 \sigma^{3/2}$ for small $\sigma$,  where the power $3/2$ was first derived by Khlopov and Polnarev \cite{PolnarevKhlopov1981}. This approximation is valid up to $\sigma \sim 0.05$. 
For $\sigma \sim \mathcal O(1)$,  $\beta_{inhom}^{caus}$ is approximated by using the error function of \eq{large-beta}. 
On the other hand, for $\sigma \ll1$, $\beta_{inhom}^{inst}$ behaves as \eq{small-beta-without} with the power $3/2$ in the same way as $\beta_{inhom}^{caus}$. 
However, coefficients are different. $\beta_{inhom}^{caus}$ is larger by about an order of magnitude than $\beta_{inhom}^{inst}$.

Before concluding this article, we emphasize that our formula for formation probability indeed evaluates the minimum rate for PBH formation because of the following reason:
In this work, we derived our formula  \eq{improved-criterion}  from the condition \eq{PBH-condition-revised} assuming conservatively that information of singularity formation propagates at the maximum possible velocity, i.e., the speed of light.
However, it is expected that the smaller the propagation speed of information, the more PBHs tend to be formed. The slow propagation is indeed possible, e.g., sound speed of a fluid if we assume that the dust is composed of particles with small but nonvanishing velocity dispersion in reality. 
Even if the signal from the central singularity propagates to the observer at the surface before she or he becomes trapped in our model, it does not necessarily mean that black hole formation is prohibited in the corresponding realistic collapse. 
In fact, it is also likely that the effect of pressure gradient in the vicinity of the center just slows down the rise of the central density, while the evolution of the surrounding 
region proceeds as in our model so that the surrounding mass is accreted by
the central region. 
In such a case, the collapse can still lead to black hole formation rather than naked singularity formation (or dispersion of dust particles) if the mass accretion results in a very compact central mass as discussed in Ref.~\cite{HaradaYooKohriNakaoJhingan2016}.

In Sec.~\ref{section-inhomo-aniso} we have derived the probability including the effects of inhomogeneity and anisotropy, though it may be still complicated to take the effect of the spin into account. 
We leave analysis of the probability including the spin effect for future work.

\section*{Acknowlegements}
T. Kokubu thanks Ken-ichi Nakao, Takahiro Terada and Takahiko Matsubara for fruitful advices.
This work is supported in part by JSPS KAKENHI grant No.~JP16H06342 (K.Kyutoku), No.~JP17H01131 (K.Kohri and K.Kyutoku), No.~JP18H04595 (K.Kyutoku), and MEXT KAKENHI Grant Nos.~JP15H05889, JP18H04594 (K.Kohri).
This work was supported by JSPS Leading Initiative for Excellent Young Researchers.

\end{document}